\documentstyle[preprint,eqsecnum,aps]{revtex}

\thicklines
\tightenlines
\begin{document}
\draft
\title{\bf UNIVERSALITY FOR SU(2) YANG-MILLS THEORY IN (2+1)D}
\author{C.J. Hamer\cite{byline1}, M. Sheppeard\cite{byline2}
and Zheng Weihong\cite{byline3} }
\address{School of Physics,    \\
The University of New South Wales,        \\
Sydney, NSW 2052, Australia.}
\author{and D. Sch\"utte\cite{byline4}}
\address{Institut f\"ur Theoretiche Kernphysik
der Universit\"at Bonn,\\
Nussallee 14-16, 53115 Bonn, Germany.}
\date{Nov. 21, 1995}
\maketitle
\begin{abstract}
The Green's Function Monte Carlo method of Chin et al is applied
to SU(2) Yang-Mills theory in (2+1)D. Accurate measurements are
obtained for the ground-state energy and
mean plaquette value, and for various Wilson loops. The
results are compared with series expansions and coupled
cluster estimates, and with the Euclidean Monte Carlo
results of Teper. A striking demonstration of
universality between the Hamiltonian
and Euclidean formulations is obtained.
\end{abstract}
\pacs{PACS Indices:  11.15.Ha; 12.38.Gc \\ \\ \\
submitted to Phys. Rev. D, DL5695}
% \newpage

%\narrowtext
\section{INTRODUCTION}
The Euclidean Monte Carlo approach is now well entrenched as
the most accurate and reliable method of studying lattice
gauge theory in the weak-coupling regime. In the
Hamiltonian formulation, however,
the development of Monte Carlo methods has been less rapid, and
much work remains to be done.

Most of the early attempts at Monte Carlo calculations in the
Hamiltonian formulation\cite{pat,dah,koo,ele,yun} were
studies of U(1) Yang-Mills theory, using a strong-coupling
(electric field) representation, which gives rise to a discrete set
of basis states. Unfortunately, such techniques appear to fail
for non-Abelian theories, because the introduction of
Clebsch-Gordan coefficients leads to the infamous ``minus sign"
problem. That is, the Hamiltonian in this representation gives
rise to amplitudes of both positive and negative signs contributing
to the same final state, which interfere with each other.
Monte Carlo  techniques are not able to recognize and cancel these
intefering amplitudes, and consequently they build up
as ``noise" in the calculation.
One needs to develop methods based on  a weak-coupling
representation, which should be able to avoid this problem.

A weak-coupling algorithm has in fact been introduced
by Heys and Stump\cite{Hey83} and Chin, Negele and Koonin\cite{Chi84},
based on the Green's Function Monte Carlo (GFMC) techniques of
Kalos and collaborators\cite{Cep79,kat}. We have previously
used an improved version of this technique to perform an
accurate finite-size scaling study of the U(1) gauge model in
(2+1) dimensions\cite{Ham94}. Here we report on an
application of the same technique to a non-Abelian model,
SU(2) Yang-Mills theory in (2+1)D.

A theoretical discussion of the SU(2) model has been given by
Feynman\cite{Fey81}, who argued that any correlations would be
of finite range, and that the flux between external charges is
restricted to a ``tube" of finite extent, leading to
a strongly confining
linear potential. All physical quantities should simply scale
according to their physical dimensions in the continuum limit,
so that the mass gap behaves as
\begin{equation}
M a \sim c_1 g^2 \quad {\rm as} \quad a \to 0
\end{equation}
and the string tension behaves as
\begin{equation}
\sigma a^2 \sim c_2 g^4  \quad {\rm as} \quad a \to 0
\end{equation}
where $a$ is the lattice spacing, and $g^2 = e^2 a$ is the
dimensionless coupling.

Numerical treatments have borne out these expectations very
well, including both variational\cite{ari,sur,Hof83,ari2}
and numerical\cite{dun,mun,hok,irb,far,Ham85,Ham92}
calculations. In the Hamiltonian formulation, most studies have
employed some form of linked cluster expansion, such as
strong-coupling series expansions\cite{Ham85,Ham92}, coupled-cluster
expansions\cite{guo,lle}, and the so-called\cite{Ham85}
``Exact Linked-Cluster Expansion" (ELCE).
One of our major aims is to make a comparison between
the results of
the GFMC method and these linked-cluster expansion methods.

In the Euclidean formulation, Teper\cite{Tep92} has recently
applied the full power of modern  Monte Carlo techniques
to this model.
His results are very accurate, and now provide the benchmark
with which all other approaches must be compared at weak coupling.

In Section II and III of the paper we briefly summarize the Monte Carlo
method and the coupled-cluster method to be employed, and
in Section IV we present the results. The GFMC results are extremely
good for the ground-state energy and mean plaquette value, but
not as good as the linked-cluster estimates for excited-state
properties such as the string tension.
A comparison with Teper's results\cite{Tep92} gives striking evidence
of universality between the Hamiltonian and Euclidean formulations,
not only in the continuum limit, but even at finite
couplings.  Further discussion is given in
Section V.

%\newpage
\section{Monte Carlo Method}

We have employed a Green's Function Monte Carlo approach to SU(2)
Hamiltonian lattice gauge theory which was discussed previously
by Chin et al\cite{Chi85}, and we shall merely summarize some of the
key points here. The lattice Hamiltonian is given by
\begin{equation}
H = {g^2\over 2a} \{ \sum_l E_l^a E_l^a -
\lambda \sum_p {\rm Tr} U_p \} \label{eq3}
\end{equation}
where $E_l^a$ is a component of the electric field at link $l$,
$\lambda = 4/g^4$, and $U_p$ denotes the product of four
link operators $U_l$ around an elementary plaquette. The
commutation relation between electric field and link operators
at each link may be taken as:
\begin{equation}
[ E_l^a , U_l] = {1\over 2} \tau^a U_l ~,
\end{equation}
choosing the $E_l^a$ as left generators of SU(2). For
calculational purposes, it is most convenient to ``scale out"
a factor $g^2/a$, and work with the dimensionless Hamiltonian
\begin{equation}
H = {1\over 2} \sum_l E^a_l E^a_l -
{\lambda \over 2} \sum {\rm Tr} U_p~. \label{heq}
\end{equation}
This will be assumed henceforth, unless stated otherwise.

The Green's Function Monte Carlo technique employs the operator
$\exp ( - \tau (H-E))$ as a projector onto the ground state
$\vert \Psi_0 \rangle$:
\begin{equation}
\vert \Psi_0 \rangle \propto \lim_{\tau \to \infty} e^{-\tau(H-E) }
\vert \Phi \rangle \label{evo}
\end{equation}
where $\vert \Phi \rangle$ is any suitable trial state.
To procure some variational guidance, one performs a ``similarity
transformation" with the trial wave function $\Phi$, and evolves
the product $\Phi \vert \Psi_0\rangle$ in imaginary time.
The heart of the procedure is the calculation of the matrix element
corresponding to a single small time step $\Delta \tau$:
Chin et al\cite{Chi85} show that
\begin{eqnarray}
\langle {\bf x}' & \vert & \Phi e^{-\Delta \tau (H-E)} \Phi^{-1} \vert
{\bf x} \rangle \nonumber \\
&=& \prod_l \langle U_l' \vert N \{ \exp (-{1\over 2} \Delta \tau
E_l^a E_l^a ) \exp [ \Delta \tau E_l^a (E_l^a
\ln \Phi )] \} \vert U_l \rangle \nonumber \\
&&  \exp \{ \Delta \tau [ E - \Phi^{-1} H \Phi (x) ]\}
+ O(\Delta \tau^2 ) \nonumber \\
&& \equiv p({\bf x}', {\bf x} ) w (x) + O(\Delta \tau^2 )
\end{eqnarray}
where ${\bf x} = \{ U_l \}$ denotes an entire lattice configuration
of link fields.

In the Monte Carlo procedure, the product
$\Phi \vert \Psi \rangle $ is simulated by the density of
an ensemble of random walkers in configuration space. At the
$k$th time step, the ``weight" of each walker at ${\bf x}_k^i$
is multiplied by $w ({\bf x}_k^i )$, and the next ensemble
$\{ {\bf x}_{k+1}^i \}$ is evolved from $\{ {\bf x}_k^i \} $ according
to the matrix element $p({\bf x}_{k+1}, {\bf x}_k )$.
Chin {\it et al.} \cite{Chi85} show that the effect of
$p({\bf x}_{k+1}, {\bf x}_k )$ is to alter each link variable $U$ in
$\{ {\bf x}_k \}$ to $U'$ by a Gaussian random walk plus
a ``drift step" guided by the trial function:
\begin{equation}
U' = \Delta UU_d U
\end{equation}
where $U_d =\exp [ i {1\over 2} \tau^a (i\Delta \tau E^a \ln \Phi ) ] $
is the drift step, and $\Delta U$ is an SU(2) group element Gaussian
distributed in distance from the identity with zero mean and
variance $\langle \Delta s^2 \rangle = 3 \Delta t$.

At the end of each iteration or ``time step"
$\Delta \tau$, the trial energy $E$ is adjusted to compensate
for any change in the total weight of all walkers in the
ensemble; so that at equilibrium its average value is equal to
the ground-state energy $E_0$. A ``branching" process is also carried
out: any walker whose weight has gone above
(say) 2 is split into two new walkers, while any two walkers
with weights less than (say) $1/2$ are combined into one,
chosen randomly according to weight from the originals.
This procedure of ``Runge smoothing"\cite{Run92}
maximizes statistical accuracy by keeping the weights of
all the walkers within fixed bounds, while minimizing any
fluctuations in the total weight due to the branching
process. Various corrections due to the finite time interval
$\Delta \tau $ have been ignored in this discussion, and the limit
$\Delta \tau \to 0$ must  be taken in some fashion to eliminate
such corrections.

The link variables are elements of the group SU(2), and
are most conveniently represented by quaternions $x^\mu =
(x^0, x^a)$
\begin{equation}
U = \exp (-i {1\over 2} \tau^a A^a )
= \cos ({1\over 2} \rho ) - i \tau^a \hat{n}^a
 \sin ({1\over 2} \rho ) \equiv x^0 - i \tau^a x^a
\end{equation}
where $\{a=1,2,3\}$, $\rho^2 = A^a A^a, \hat{n}^a = A^a/\rho$, and
\begin{equation}
x^0 x^0 + x^a x^a =1
\end{equation}
Note that then
\begin{equation}
{\rm Tr} (U) = 2 x^0~.
\end{equation}

The product of two link variables is then easily
found by quaternion multiplication:
\begin{equation}
U(x) U(y) = (x^0 y^0 - x^a y^a ) - i \tau^a ( x^0 y^a +
y^0 x^a + \epsilon^{abc} x^b y^c )
\end{equation}

To generate the Gaussian-distributed element $\Delta U$ near
the identity, we\cite{Chi85} simply generate 3 random numbers
$A^1$, $A^2$, $A^3$ from a Gaussian distribution with zero mean
and variance $\Delta \tau $, and set
\begin{equation}
x^\mu = ( \cos({1\over 2} \rho ) , {A^a\over \rho}
\sin({1\over 2} \rho ))
\end{equation}
where $\rho^2 = A^a A^a$.

The trial wave function is chosen to be the
one-parameter form\cite{Chi85}:
\begin{equation}
\Phi = \prod_{p } \exp [ {\alpha \over 2} {\rm Tr} U_p ]
\end{equation}
Then the drift step is
\begin{eqnarray}
U_d &=& \exp [ i {1\over 2} \tau^a
    (i \Delta \tau E^a \ln \Phi )] \nonumber \\
& =& \exp ( - i {1\over 2} \tau^a A^a_l )
\end{eqnarray}
where
\begin{equation}
A_l^a = - {1\over 2} \alpha \Delta \tau \sum_{p\in l} x^a_{p,l}
\end{equation}
and the sum runs over plaquettes adjacent to the
link $l$, with the $x_{p,l}^a$
being ``rotated"quaternion elements as defined in Appendix A of
Chin {\it et al.}\cite{Chi85}. Finally, the trial
function energy factor is
\begin{equation}
\Phi^{-1} H \Phi = \sum_l \sum_{p\in l} \{ -
{\alpha^2 \over 8} x_{p,l}^a x_{p,l}^a
+ {1\over 4} (3\alpha/2 - \lambda ) x_p^0 \}
\end{equation}
where $x_p^0$ is the zeroth quaternion element of the
plaquette operator or group element $U_p$, and the sum $p\in l$ runs
over the two plaquettes adjacent to the link $l$.

The expectation value of an observable $Q$ in the
ground state can be measured
by the technique of Hamer {\it et al.} \cite{Ham94}. If one
adds a small perturbation $yQ$ to the Hamiltonian
\begin{equation}
H' = H + y Q
\end{equation}
then by the Feynman-Hellmann theorem the required expection
value is geven by
\begin{equation}
\langle Q \rangle_0 = {d E'_0 \over d y}{\Big \vert}_{y=0}
\end{equation}
Perform a Taylor expansion in $y$ for the eigenvector and eigenvalue
\begin{eqnarray}
\vert \Psi_0 (y,\tau )\rangle &=&
\vert \Psi_0^0 (\tau) \rangle + y \vert \Psi_0^1 (\tau)\rangle
+ O(y^2 )  \\
E'_0 (y) &=& E_0^0 + y E_0^1 + O(y^2 )
\end{eqnarray}
substitute in the evolution equation (\ref{evo})
and equate powers of $y$ to obtain:
\begin{equation}
\vert \Psi_0^0 (\tau + \Delta \tau ) \rangle
 = e^{-\Delta \tau (H-E) }
\vert \Psi_0^0 (\tau )\rangle \label{eq21}
\end{equation}
(which is the same as in the unperturbed case), and
\begin{equation}
\vert \Psi_0^1 (\tau + \Delta \tau ) \rangle =
e^{-\Delta \tau (H-E) } \vert \Psi_0^1 (\tau )
\rangle + \Delta \tau (E_0^1 - Q)
\vert \Psi_0^0 (\tau) \rangle  \label{eq22}
\end{equation}
Equation (\ref{eq22}) is an evolution equation for
$\vert \Psi_0^1 \rangle $ of similar structure to
(\ref{eq21}). It is simulated by giving a ``secondary"
weight to each walker in the ensemble to simulate
$\vert \Psi_0^1 \rangle $, and evolving it according to
(\ref{eq22}). The value of $E_0^1$ is adjusted to keep
the total of all secondary weights constant after each iteration.
At equilibrium, its average value gives an estimate of
$\langle Q \rangle_0 $.

% We need something about the coupled cluster algorithm.
% Outline of coupled cluster algorithm.

\section{Coupled Cluster Expansion Method}

The coupled cluster expansion method has been extensively
used in many-body theory, and has recently been introduced to
lattice gauge theory by Bishop\cite{bis}, Llewellyn-Smith
 and Watson\cite{lle}, and Guo {\it et al}\cite{guo},
although the truncation schemes used there are different.
The basic idea of this expansion is to assume the ground
state $\vert \Psi_0 \rangle$ and first excited state
(the glueball wavefunction)
$\vert \Psi \rangle$ of the Hamiltonian in
eq. (\ref{eq3}) can be represented by
an exponential form:
\begin{eqnarray}
\vert \Psi_0 \rangle &=& e^{R(U)} \vert 0 \rangle
 \nonumber \\  && \\
\vert \Psi   \rangle &=& F(U)  e^{R(U)} \vert 0
\rangle \nonumber
\end{eqnarray}
where $R(U)$ and $F(U)$ are functions of loop variables
 and the state $\vert 0 \rangle $ is the strong-coupling
 ground-state,  defined by
\begin{equation}
E_l^a \vert 0 \rangle =0~.
\end{equation}
The eigenvalue equation for $H$ can then be written as:
\begin{eqnarray}
\sum_l ([ E_l, [ E_l, R]] && + [E_l, R] [E_l, R] ) -
{4\over g^4} \sum_p {\rm Tr} (U_p) = {2 a\over g^2}
\epsilon_0 \nonumber \\
 && \label{eq33c} \\
\sum_l ([ E_l, [ E_l, F]] && + 2 [E_l, F] [E_l, R] ) =
{2 a\over g^2} (\epsilon_1 - \epsilon_0) \nonumber
\end{eqnarray}
where $\epsilon_0$ ($\epsilon_1$) is the ground state(the
first excited state) energy. $R(U)$ and $F(U)$ can be
 decomposed according to the order of graphs,
\begin{eqnarray}
R &=& \sum_i R_i \nonumber \\
 && \\
F &=& \sum_i F_i \nonumber
\end{eqnarray}
and the lowest order
term of $R$ and $F$ is:
\begin{eqnarray}
% R_1 &=& c_1 ~ \grOne  \nonumber \\
R_1 &=& c_1 ~
  \begin{picture}(10,10)
  \put(0,0){\framebox(10,10){}}
  \end{picture}
     \nonumber \\
 &&  \\
% F_1 &=& f_1 ~ \grOne  \nonumber
F_1 &=& f_1 ~
   \begin{picture}(10,10)
   \put(0,0){\framebox(10,10){}}
   \end{picture}
   \nonumber
\end{eqnarray}
The graphs of order $i$ are generated by
\begin{equation}
\sum_{j=1}^{i-1} [E_l, R_j] [E_l, R_{i-j} ]
\end{equation}
in equation (\ref{eq33c}).
In order to make the calculation possible, some truncation
scheme to truncate the eigenvalue equation must be used.
The truncation scheme used by Llewellyn-Smith and Watson\cite{lle} is
\begin{eqnarray}
\sum_l ([ E_l, [ E_l, \sum_{i=1}^n R_i]] && + \sum_{i,j=1}^n
[E_l, R_i] [E_l, R_j] ) -
{4\over g^4} \sum_p {\rm Tr} (U_p) =
{2 a\over g^2} \epsilon_0  \nonumber \\
 &&     \label{schlle}  \\
\sum_l ([ E_l, [ E_l, \sum_{i=1}^n F_i]] && + 2 \sum_{i,j=1}^n
[E_l, F_i] [E_l, R_j] )  = {2 a\over g^2}
(\epsilon_1 - \epsilon_0) \nonumber
\end{eqnarray}
where the new graphs generated by $[E_l, R_i] [E_l, R_j]$ and
$[E_l, F_i] [E_l, R_j]$
are simply discarded. Guo {\it et al}\cite{guo} have argued that
because of this, the continuum limit of this system could not
be preserved, and they have proposed a better truncation scheme:
\begin{eqnarray}
\sum_l ([ E_l, [ E_l, \sum_{i=1}^n R_i]] && + \sum_{i+j\le n}
[E_l, R_i] [E_l, R_j] ) -
{4\over g^4} \sum_p {\rm Tr} (U_p) =
 {2 a\over g^2} \epsilon_0 \nonumber \\
 &&    \label{schguo}   \\
\sum_l ([ E_l, [ E_l, \sum_{i=1}^n F_i]] && + 2 \sum_{i+j\le n}
[E_l, F_i] [E_l, R_j] )  = {2 a\over g^2}
 (\epsilon_1 - \epsilon_0) \nonumber
\end{eqnarray}

The most tedious task for the high-order approximations
is to generate a list of  independent loop
 configurations and derive the
nonlinear coupled equations.
So far, all these calculations in lattice gauge theory have been
carried out by hand. We have tried to develop  computer
algorithms to overcome this problem.
Borrowing some ideas from our computer algorithms used to
generate a list of clusters for our linked-cluster
series expansions and $t$-expansions\cite{texp},
a preliminary program was developed.
Up to fourth order, a list of 70 graphs was generated,
whereas Llewellyn-Smith and Watson\cite{lle}
only obtained 69 graphs by hand.
Some results from the truncation scheme (\ref{schguo})
were presented in a previous paper\cite{guo3}.
Here we make a comparison of the results of
different truncation schemes with the GFMC simulation.

\section{RESULTS}
The GFMC method was used to calculate the
 ground-state energy, mean
plaquette value, and some Wilson loop expectation values
for lattice sizes of $2\times 2$ up to $6\times 6$ sites,
all with periodic boundary conditions. At each coupling
$\lambda$, the variational parameter $\alpha$ was
adjusted to minimize the error by a series of trial runs.
Production runs typically employed an ensemble size of
2000 walkers for 50,000 iterations, with step sizes
$\Delta \tau = 0.05 $ and 0.01, followed by a linear extrapolation
to $\Delta \tau =0$.
The first  8K iterations were discarded  to allow for equilibration,
and the results were averaged over blocks of 5K iterations
before estimating the error to minimize correlation effects.

\vskip 1pc
\noindent {\bf Ground-State Energy}

A table of estimates of the ground-state energy per site is given
in Table I.
The finite-size dependence of these results is illustrated in Figure 1.
It can be seen that in the strong-coupling region (small $\lambda$)
the convergence to the bulk limit is very rapid. At weak coupling
(large $\lambda$), however, the convergence is slower, as one
might expect, and at $\lambda =16$ the data can be quite well fitted
by a straight line in $1/L^3$ (where $L$
  is the lattice size). The straight
line fit in Fig. 1c corresponds to
\begin{equation}
\omega_0(L) = - 10.81 - 1.19/L^3 \label{eq23}
\end{equation}
This behaviour appears similar to that of the U(1) model\cite{Ham94},
and we shall discuss it further in the concluding section.

It can be seen from Table I that the estimated
 bulk limits are in excellent
agreement with earlier strong-coupling series
 estimates\cite{Ham92}, within
errors, which provides a check that our algorithm is working
correctly. It is also noteworthy that in the weak-coupling region
beyond $\lambda =4$ the Monte Carlo results are much more accurate
than the series estimates.

The data in the weak-coupling region can be compared with the
asymptotic weak-coupling expansion obtained by Hofs\"{a}ss and
Horsley\cite{Hof83}:
\begin{equation}
\omega _0 \sim - \lambda + 1.4372 \lambda^{1/2} \quad {\rm as} \quad
\lambda \to \infty \label{eq24}
\end{equation}
Figure 2 shows that the Monte Carlo results appear to be
in excellent agreement with this prediction as the continuum
limit is approached.
The fit to the data shown in Figure 2 corresponds to
\begin{equation}
\omega_0 = -\lambda + 1.4372 \lambda^{1/2} - 0.59(2)
\end{equation}

\vskip 1pc
\noindent {\bf Mean Plaquette Value}

The mean plaquette value
\begin{equation}
P = {1\over 2} \langle {\rm Tr} U_p \rangle_0
= - {1\over N} {d \omega_0 \over d \lambda}
\end{equation}
by the Feynman-Hellmann theorem. This
expectation value has also been
estimated by the methods outlined in the previous Section.
The results are listed in Table 2: again,
 it can be seen that the Monte Carlo
estimates of the bulk limit are in excellent agreement with
previous strong-coupling series estimates\cite{Ham92}. The Monte
Carlo results are again more accurate than the series
estimates beyond $\lambda =4$.

The behaviour in the weak-coupling region
is illustrated in Figure 3.
The data are described almost exactly by
the asymptotic prediction
obtained from (\ref{eq24}):
\begin{equation}
1- P \sim  0.7186 \lambda^{-1/2} \quad {\rm as}
\quad {\lambda \to \infty} \label{eq27}
\end{equation}

\vskip 1pc
\noindent {\bf String Tension}

In attempting to estimate the string tension, we have
used the method of Wilson loops. The Wilson loop
expectation values, defined as
\begin{equation}
W(I,J) = {1\over 2} \langle {\rm Tr} U(I,J) \rangle_0
\end{equation}
where $U(I,J)$ denotes a product of
 link variables around a rectangular
loop of $I\times J$ links, were computed using the method of
``secondary weights" outlined in the previous section.
These can be combined to form the Creutz ratios:
\begin{eqnarray}
\chi (I,J) &=& - \ln {\Big [} {W(I,J) W(I-1, J-1) \over
W(I,J-1) W(I-1,J) } {\Big ]} \nonumber \\
 && \to a^2 \sigma  \quad {\rm as} \quad I,J \to \infty
\end{eqnarray}
where $\sigma$ is the string tension. Table 2 shows
some typical Wilson loop estimates. Some problems are
immediately apparent. At small $\lambda$, the estimates
decrease extremely rapidly with the size of the loop; while
at large $\lambda$,  they decrease very slowly, and larger
lattices are really needed to get reliable values for
the string tension. In both cases, the resulting error
in the string tension estimate is very large.

Figure 4 shows the GFMC estimates of the string
tension, obtained from $\chi (2,2)$ as measured  on
the $6\times 6$ lattice, together with some earlier Hamiltonian
estimates by Hamer and Irving\cite{Ham85} obtained using
the ELCE method, and the Euclidean results of Teper
\cite{Tep92}. A number of points may be noted:

i) The Euclidean MC results are an order of magnitude more
accurate than any Hamiltonian estimates to date,
 and extend to weaker couplings;

ii) The GFMC results are relatively inaccurate and appear to
 diverge at weak couplings. We take this to indicate that both
the lattice size and the size of the Wilson loops are
too small.  Indeed, the $\chi (3,3)$ estimates appear
somewhat lower than $\chi (2,2)$, but are
so inaccurate as to be not worth plotting.
The GFMC and ELCE results do not agree; but this is expected,
because the first is a `spacelike' (correlation length)
estimate,  while the second is a `timelike' (energy)
estimate, and the two differ by a factor equal to the
`speed of light'.

iii) On the positive side, it can be
 seen that the ELCE estimates
for the string tension approach a limit very similar to
Teper's results\cite{Tep92} as $g\to 0$. This provides
nice evidence of universality between
 the Euclidean and Hamiltonian
formulations in the continuum limit.

Let us expand on the last point  a little further.
When $g$ is non-zero, the Euclidean and Hamiltonian
results are not directly comparable because there is a
difference in scale (i.e. coupling) between the two
formulations, and the speed of light in the Hamiltonian
is not equal to unity. In four dimensions,
the relationship between the two scales was
 calculated long ago to one-loop
order in weak-coupling perturbation theory by
Hasenfratz and Hasenfratz\cite{Has81}. The calculation has
recently been repeated for the three-dimensional case
by one of us\cite{Ham95}.
The results for this model may be summarized as:
\begin{equation}
{1\over g_H^2 } = {1\over g_E^2} -0.01924 + O(g_E^2) \label{eq30}
\end{equation}
where $g_H$, $g_E$ are the couplings in the Hamiltonian
and Euclidean models, respectively; and
\begin{equation}
c = 1-0.08365 g_E^2 + O(g_E^4) \label{eq31}
\end{equation}
for the `speed of light', with the Hamiltonian normalized
as in equation (\ref{eq3}). To make a direct comparison between
string tension estimates, the `timelike' ELCE estimates must be
divided by a factor of $c$, and shifted from coupling $g_H$
to the equivalent $g_E$, given by (\ref{eq30}). The revised
estimates are then given by
\begin{equation}
{4 a \sigma_E^{1/2} \over g_E^2 } =
{4 a \sigma_H^{1/2} \over g_H^2}
[ 1+ 0.0616 g_E^2 ]
\end{equation}
and are compared with Teper's results in Figure 5.
 It can be seen that the two
sets of estimates now lie almost on the same
 curve, and the remaining
discrepancy (of order 3\%) may easily be attributed to higher-order
(two loop) corrections. This provides even stronger evidence of
universality between the two formulations.

\vskip 1pc
\noindent {\bf Mass Gap}

It is much more difficult to measure the energy of an excited
state than it is for the ground state using
 the GFMC method\cite{Ham94},
and we have  no GFMC results to report for the mass gap.
There have been a number of Hamiltonian estimates of the mass gap
using various linked cluster expansion techniques\cite{guo,lle},
however, and it is interesting to compare these with the Euclidean
results of Teper\cite{Tep92}.

Let us begin with estimates of the dimensionless ratio
$M/\sqrt{\sigma}$. Figure 6 compares the results obtained
 by Hamer and Irving\cite{Ham85} with the Euclidean results
of Teper\cite{Tep92}. The Hamiltonian estimates have been slightly
shifted from coupling $g_H^2$ to the equivalent coupling $g^2_E$
using equation (\ref{eq30}), and `renormalized'
by the speed of light $c$ according to
\begin{equation}
{M_E\over \sqrt{\sigma_E} } = {M_H \over \sqrt{\sigma_H c} }
= {M_H \over \sqrt{\sigma_H } } (1+0.0418 g_E^2)
\end{equation}
It  can be seen that the two sets of data
match rather well: once more, in excellent agreement with
universality.

Next, Figure 7 shows a number of different Hamiltonian
estimates of the mass gap itself, or more precisely of the
quantity $Ma/g^2$, which should approach
 a finite value in the continuum
limit. The strong-coupling series expansion estimates\cite{Ham92}
rise fairly abruptly around $\beta = 4/g^2\simeq 3$, and then
begin to level off towards an asymptotic value estimated
previously\cite{Ham92} as 2.22(5). The higher-order
coupled-cluster estimates by the truncation schemes used by
 Llewellyn-Smith and Watson\cite{lle}
and Guo {\it et al.}\cite{guo} also show
 a rise until $\beta\simeq 4$,
but their behaviour beyond that point is somewhat variable.
Teper\cite{Tep92} does not give direct
 estimates of the mass gap itself;
but using his results for $\sigma$ and the ratio $M/\sqrt{\sigma}$,
together with equations (\ref{eq30}) and (\ref{eq31}), one may infer
a behaviour for the Hamiltonian mass gap at weak coupling
which is shown as a heavy solid line in Fig. 7.
It would appear that $Ma/g^2$ actually reaches a peak at around
$\beta \simeq 4$, and then declines slowly towards the asymptotic
 value 1.59(2). The strong-coupling series extrapolations were unable
to pick up this decrease at weak coupling. Some of the
coupled-cluster approximants perhaps give
some indication of it, but not in a very consistent or reliable
manner.

% \newpage

\section{DISCUSSION}
The most interesting result of this paper is the
remarkable demonstration of universality between the
Euclidean and Hamiltonian formulations
 which has been obtained.
If account is taken of the difference in scale between the
two formulations as predicted in one-loop perturbation
theory\cite{Ham95}, then the Euclidean
 results of Teper\cite{Tep92}
and the Hamiltonian results of Hamer and Irving\cite{Ham85} for
the string tension and the ratio $M/\sqrt{\sigma}$ fall
almost on top of each other, not only in the continuum
limit, but over a whole range of weak couplings. This provides
pleasing confirmation of the hypothesis of universality.
There is little doubt that the hypothesis is correct, but
it is important to check it wherever possible, since
it underpins the whole program of lattice gauge
theory.

One of our objectives has been to continue the development of
Monte Carlo techniques for Hamiltonian lattice gauge theory,
which currently lag behind Euclidean techniques
by about a decade. We have shown that the GFMC method of
Chin {\it et al.} \cite{Chi84,Chi85}, when improved by
``Runge smoothing"\cite{Run92}, is capable of giving an accurate
description of the ground-state of this
 non-Abelian model deep into the
weak-coupling regime. The ground-state
 energy has been obtained to an
accuracy of around 2 parts in $10^4$, and the mean plaquette
to around 0.3\%. The results agree
 very nicely with both weak-coupling
and strong-coupling expansions for the model.

An interesting question concerns the finite-size scaling
behaviour to be expected in this model. For the U(1) model,
an argument based on effective Lagrangian
 theory predicts\cite{ham35} that
the ground-state energy per site scales as
\begin{equation}
\omega_0 (L) - \omega_0 (\infty ) \sim - {0.7188 c\over L^3}
\end{equation}
where $c$ is the `speed of light', based on the presence
of one massless photon degree of
 freedom per site.\footnote{There
is a delicate question of scale involved here, for which see
refs. \onlinecite{ham35,Ham94}.} This has been confirmed by
numerical calculations\cite{Ham94}.
The SU(2) model has three gauge
boson degrees of freedom per site,
and so similar arguments would predict for this model
\begin{equation}
\omega_0 (L) - \omega_0 (\infty ) \sim
 - {2.1564 c\over L^3} \label{eq33}
\end{equation}
or, with the normalization of eq. (\ref{heq}):
\begin{equation}
\omega_0 (L) - \omega_0 (\infty ) \sim
 - {4.14 \over L^3} \label{eq43}
\end{equation}
This does not agree at all well with the numerical results,
eq. (\ref{eq23}). Now in fact the mass gap
does not vanish in this model, but remains finite in the continuum
limit, so it is not unexpected that the ``zeroth order" prediction
(\ref{eq43}) should break down. It would be interesting to
carry out a higher-order calculation in weak-coupling
perturbation theory for this model to explore these questions
further. Such a calculation was carried out for the case of
free boundary conditions by M\"{u}ller and R\"{u}hl\cite{mul},
who obtained a prediction for the mass gap which was an order of
magnitude  smaller than the numerical results.

For excited state properties, the GFMC techniques have not been so
successful as yet\cite{Ham94}. An attempt was made to estimate
the string tension by means of expectation values for Wilson
loops. The results were not very accurate, and it was clear that the
lattice sizes used were too small at weak coupling. The calculation
of each Wilson loop involved the carriage of a corresponding
``secondary weight" for each walker in the ensemble, which is rather
laborious asd expensive in computer time. It seems clear that new
methods are needed here. One possibility which needs exploration
is to see if correlation functions can be estimated  by averaging
over configurations in the ensemble in some fashion, as is done in
the Euclidean framework.

A second major objective was to compare the
results of various Hamiltonian linked-cluster
expansions for this model\cite{Ham85,Ham92,guo,lle} with the
Monte Carlo results. It was found that the GFMC results
are already superior to linked-cluster estimates for the ground-state
energy and mean plaquette at weak coupling, but
are not yet competitive for excited state properties,
as discussed above.

In the case of the mass gap itself, it was possible to compare
the different linked-cluster expansion
techniques\cite{Ham92,guo,lle}. Out to $\beta \simeq 4$,
they  were found to agree with each other quite well.
Beyond that point, the convergence is poor.
 Teper's data\cite{Tep92} imply a gradual decline
in the scaled mass gap $aM/g^2$ to an asymptotic limit
of $1.59(2)$, which is not clearly indicated by any of the
expansion methods. It appears, once again, that one must
be cautious about placing too much trust in
extrapolated linked-cluster expansions deep in the weak-coupling
regime. This is not surprising, since the cluster expansions are
basically strong-coupling expansions of one or another form,
and they cannot be expected to converge very well
in the neighbourhood
of the weak-coupling (continuum)
limit, which is likely to be an essential singularity.

The linked-cluster expansion techniques have an important
role to play, nevertheless. They are presently more accurate
than the GFMC method for excited states,
 as outlined above, and will
remain useful as a check on the Monte Carlo results at
strong and intermediate couplings. They can also be used for
models containing dynamical fermions, for instance,
which at present are inaccessible by the GFMC method.

At present we are trying to improve the coupled cluster expansion
technique\cite{sch} by combining it with the D-function expansion
used in the ``ELCE'' approach of
Irving, Preece  and Hamer\cite{Irv}.
 This avoids both the use
of the Cayley Hamilton relation for the elimination of
redundancies, and also the explicit handling of
many SU(n) coupling coefficients, and allows one to define the
truncation with respect to an orthogonal basis.
 The incorporation of states in the spectrum
having arbitrary lattice momentum and lattice angular
momentum has also been possible in this framework\cite{sch}.

\acknowledgments

CJH would like to thank Dr. Russell Standish and the Sydney
Regional Center for Parallel Computing (SRCPC) at the University
of NSW for the use of their CM5 computer; and is also grateful
for a grant from the DITARD Bilateral Science and Technology
Collaboration Program which facilitated this project.

We would like to thank
Dr. Michael Teper for providing us with his
Euclidean Monte Carlo data.
This work forms part of a research project supported by a grant
from the Australian Research Council.

% \newpage

\begin{figure}
\caption{Finite-size dependence of the ground-state energy per site
$\omega_0 (L)$ at couplings $\lambda =4$, 8 and 16. The curves in
a) and b) are merely to guide the eye; the curve in c) is
a straight line.}
\label{fig1}
\end{figure}

\begin{figure}
\caption{Graph showing the weak-coupling
behaviour of the ground-state energy per site $\omega_0$.
The straight line corresponds to the asymptotic weak-coupling
prediction{\protect\cite{ari2}}, equation (\protect\ref{eq24}) }
\label{fig2}
\end{figure}

\begin{figure}
\caption{Weak-coupling behaviour of the mean plaquette value $P$.
The points are the Monte Carlo data, while the dashed line corresponds
to the asymptotic weak-coupling prediction (\protect\ref{eq27}).}
\label{fig3}
\end{figure}

\begin{figure}
\caption{The string tension $\sigma$ at weak coupling.
Circles: Euclidean MC data{\protect\cite{Tep92}}; squares:
ELCE data{\protect\cite{Ham85}}; stars: present GFMC estimates.}
\label{fig4}
\end{figure}

\begin{figure}
\caption{A comparison at weak couplings $g_E$ between
Euclidean MC{\protect\cite{Tep92}} and
ELCE{\protect\cite{Ham85}} estimates of
the string tension, including one-loop correction effects.}
\label{fig5}
\end{figure}

\begin{figure}
\caption{The ratio ${M_E / \protect\sqrt{\sigma_E } }$
plotted as a function of $\beta =4/g^2$.
Circles: Euclidean MC data \protect\cite{Tep92};
Squares: ELCE data {\protect\cite{Ham85}}. }
\label{fig6}
\end{figure}

\begin{figure}
\caption{Estimates of $aM/g^2$, where $M$ is the mass gap, as a
function of $\beta =4/g^2$. Series expansion results from
ref. \protect\onlinecite{Ham92}; coupled-cluster estimates
by the truncation scheme used by
 LLewellyn-Smith and Watson\protect\cite{lle}
and Guo {\it et al.}\protect\cite{guo};
the heavy  line is an
estimate inferred from Teper\protect\cite{Tep92}. }
\label{fig7}
\end{figure}

\widetext

\begin{table}
%\squeezetable
\setdec 0.000000
\caption{Values for the ground-state energy per site
$\omega_0 (L)$ as a function of lattice size $L$ and coupling
$\lambda$. Also listed are extrapolated estimates of the bulk limit
$L\to \infty$, together with strong-coupling
series estimates\protect\cite{Ham92}, and
 coupled-cluster estimates\protect\cite{lle}.}
\label{tab1}
\begin{tabular}{ccccccc}
{}~~~~~~~~~~~~$\lambda=$ & \multicolumn{1}{c}{0.5} &
 \multicolumn{1}{c}{1.0}
& \multicolumn{1}{c}{2.0} & \multicolumn{1}{c}{4.0}
& \multicolumn{1}{c}{8.0} & \multicolumn{1}{c}{16.0} \\
\unitlength1pt
\begin{picture}(0,0)
\put(0,21){\line(1,-1){24}}
\end{picture}
$L$  \\
\tableline \noalign{\vskip2pt}
2 &\dec $-$0.04136(2) &\dec $-$0.16154(5) &\dec $-$0.5777(1)
 &\dec $-$1.753(3)   &\dec $-$4.603(1)   &\dec $-$10.987(2)  \\
3 &\dec $-$0.04108(2) &\dec $-$0.15795(2) &\dec $-$0.5593(2)
 &\dec $-$1.7092(3)  &\dec $-$4.519(1)   &\dec $-$10.848(2)  \\
4 &\dec $-$0.04108(1) &\dec $-$0.15800(3) &\dec $-$0.5582(1)
 &\dec $-$1.7021(1)  &\dec $-$4.5048(4)  &\dec $-$10.822(1)  \\
5 &\dec $-$0.04109(1) &\dec $-$0.1590(2)  &\dec $-$0.5582(1)
 &\dec $-$1.7010(2)  &\dec $-$4.501(1)   &\dec $-$10.815(1)  \\
6 &                 &                 &\dec $-$0.5581(1)
 &\dec $-$1.7010(4)  &\dec $-$4.500(1)   &\dec $-$10.810(1)  \\
$L\to \infty$ &\dec $-$0.04109(1) &\dec $-$0.15795(3) &\dec
 $-$0.5581(1) &\dec $-$1.701(1) &\dec $-$4.499(1) &\dec $-$10.805(3) \\
SC series &\dec $-$0.04107787 &\dec $-$0.1579360  &\dec $-$0.5581(1)
  &\dec $-$1.700(5)  \\
Coupled cluster &\dec $-$0.0412(3)  &\dec $-$0.160(2)
  &\dec  $-$0.566(2)   &\dec  $-$1.70(1)   \\
\end{tabular}
\end{table}

\begin{table}
%\squeezetable
\setdec 0.000000
\caption{The mean plaquette value $P$ as a function of lattice
size $L$ and coupling
$\lambda$. Also listed are extrapolated estimates of the bulk limit
$L\to \infty$, together with strong-coupling series
estimates\protect\cite{Ham92}.}
\label{tab2}
\begin{tabular}{ccccccc}
{}~~~~~~~~~~~~$\lambda=$ & \multicolumn{1}{c}{0.5} & \multicolumn{1}{c}{1.0}
& \multicolumn{1}{c}{2.0} & \multicolumn{1}{c}{4.0}
& \multicolumn{1}{c}{8.0} & \multicolumn{1}{c}{16.0} \\
\unitlength1pt
\begin{picture}(0,0)
\put(0,21){\line(1,-1){24}}
\end{picture}
$L$  \\
\tableline \noalign{\vskip2pt}
2 &\dec 0.1649(5) &\dec  0.3116(9) &\dec 0.4986(6) &\dec  0.6505(10)
 &\dec  0.7548(7) &\dec  0.8277(8) \\
3 &\dec 0.1623(5) &\dec  0.3004(6) &\dec 0.4830(8) &\dec  0.6382(5)
 &\dec  0.7477(8) &\dec  0.8216(5) \\
4 &\dec 0.1626(5) &\dec  0.3002(5) &\dec 0.4803(4) &\dec  0.6356(8)
 &\dec  0.7448(10) &\dec  0.8197(11) \\
5 &\dec 0.1617(3) &\dec  0.2991(5) &\dec 0.4797(5) &\dec  0.6361(7)
 &\dec  0.7459(14) &\dec  0.8204(10) \\
6 &               &                &\dec 0.4816(6) &\dec  0.6357(10)
 &\dec 0.7456(12) &\dec 0.8214(21) \\
$L\to \infty$ &\dec 0.1622(5) &\dec 0.2999(5) &\dec 0.480(1)
 &\dec 0.636(1) &\dec 0.745(2) &\dec 0.820(2) \\
SC series &\dec 0.1620183 &\dec 0.3001145 &\dec 0.480918(5)
 &\dec 0.635(5) \\
\end{tabular}
\end{table}

\begin{table}
% \squeezetable
\setdec 0.000000
\caption{Sample Wilson loop estimates for the $6\times 6$ lattice}
\label{tab3}
\begin{tabular}{cccccc}
\multicolumn{1}{c}{$\lambda$ }& \multicolumn{1}{c}{$W(1,1)$}
& \multicolumn{1}{c}{$W(2,1)$} & \multicolumn{1}{c}{$W(2,2)$}
& \multicolumn{1}{c}{$W(3,2)$} & \multicolumn{1}{c}{$W(3,3)$} \\
\tableline
0.5 &\dec 0.1617(3) &\dec 0.0305(3) &\dec 0.0014(3) & $-$ & $-$ \\
2.0 &\dec 0.4816(7) &\dec 0.263(2)  &\dec 0.0932(5)
&\dec 0.034(1) &\dec 0.008(1) \\
16.0 &\dec 0.821(2) &\dec 0.709(4)  &\dec 0.569(7)
 &\dec 0.466(8) &\dec 0.361(10) \\
\end{tabular}
\end{table}

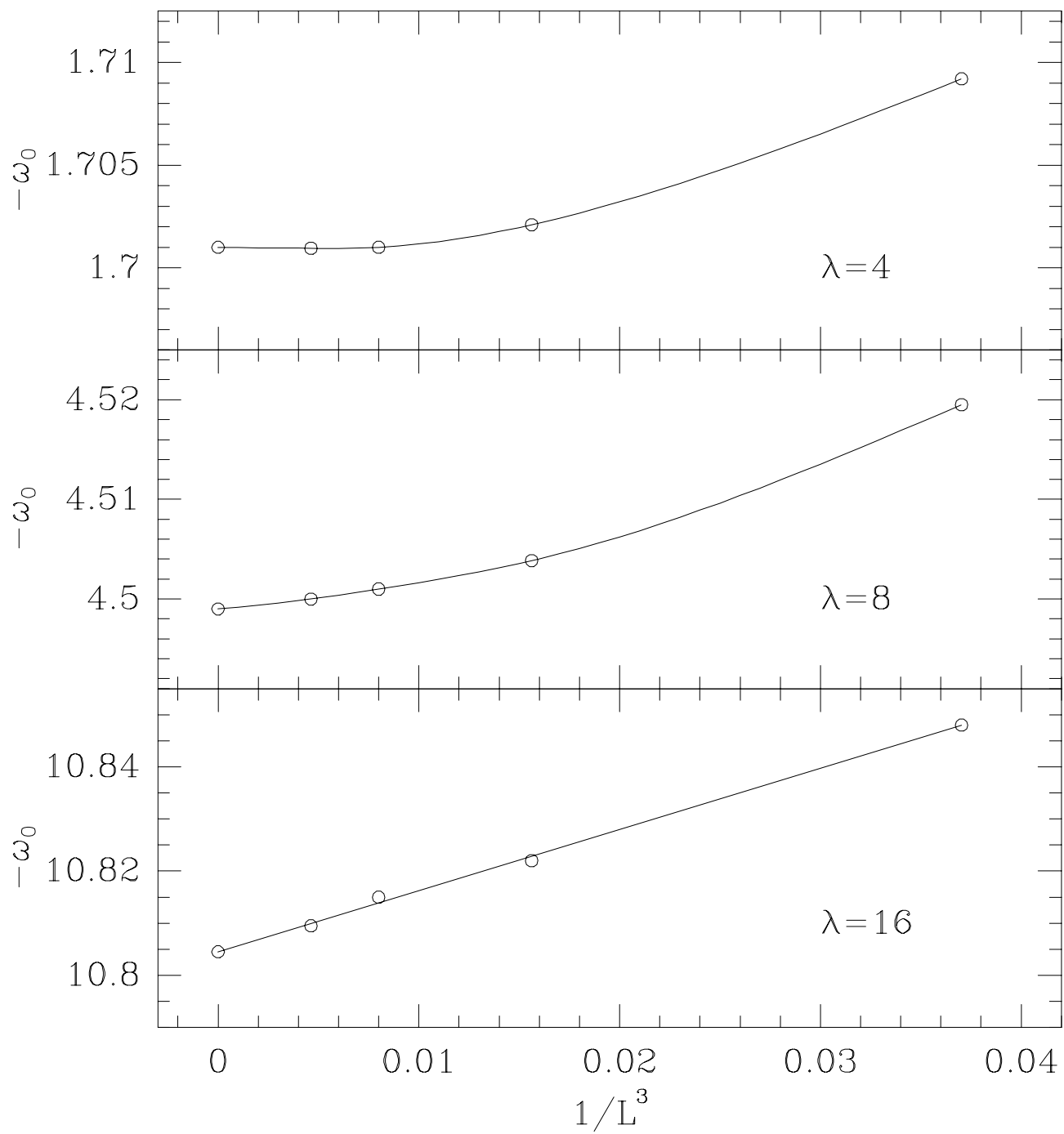
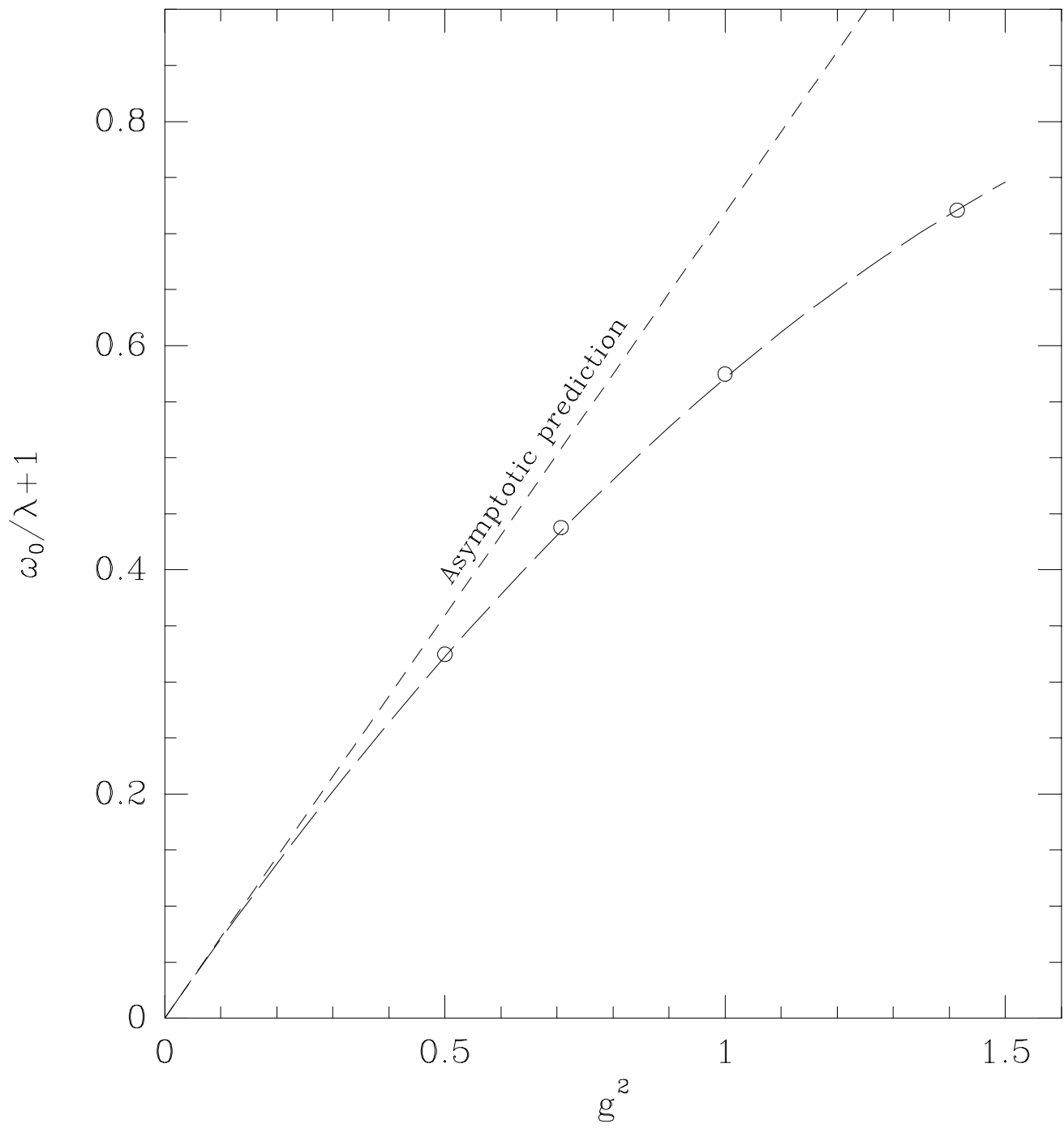
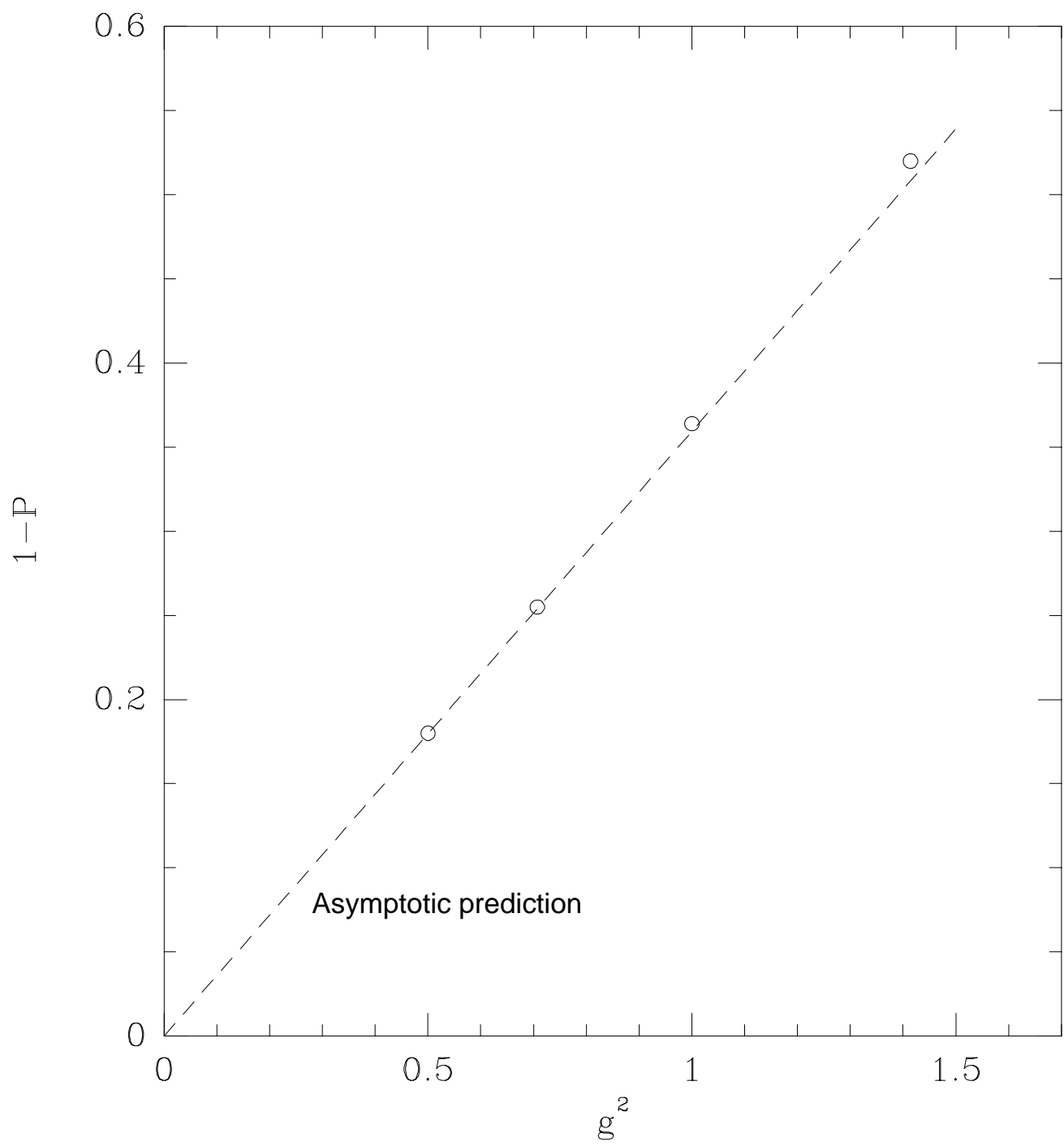
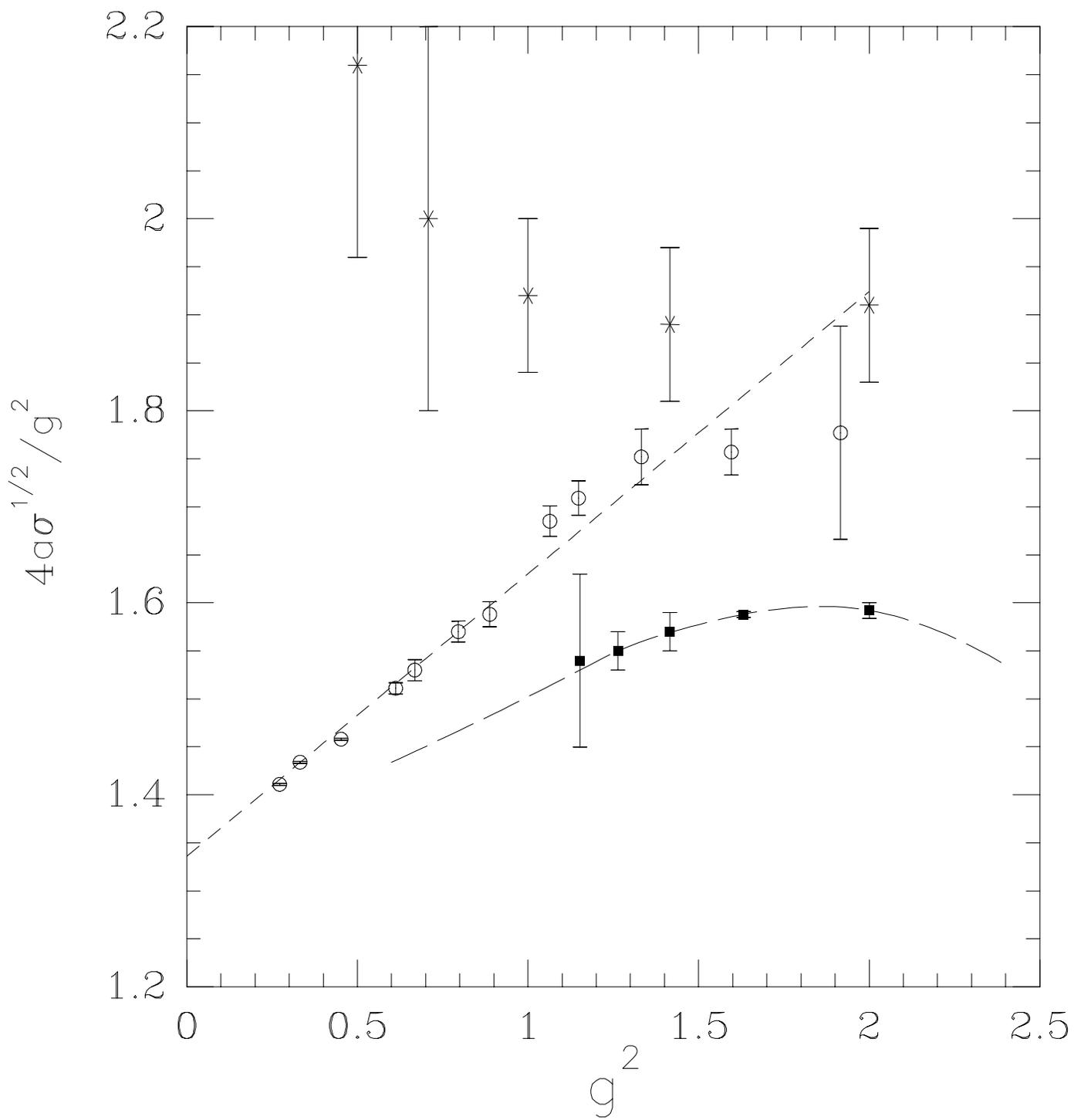
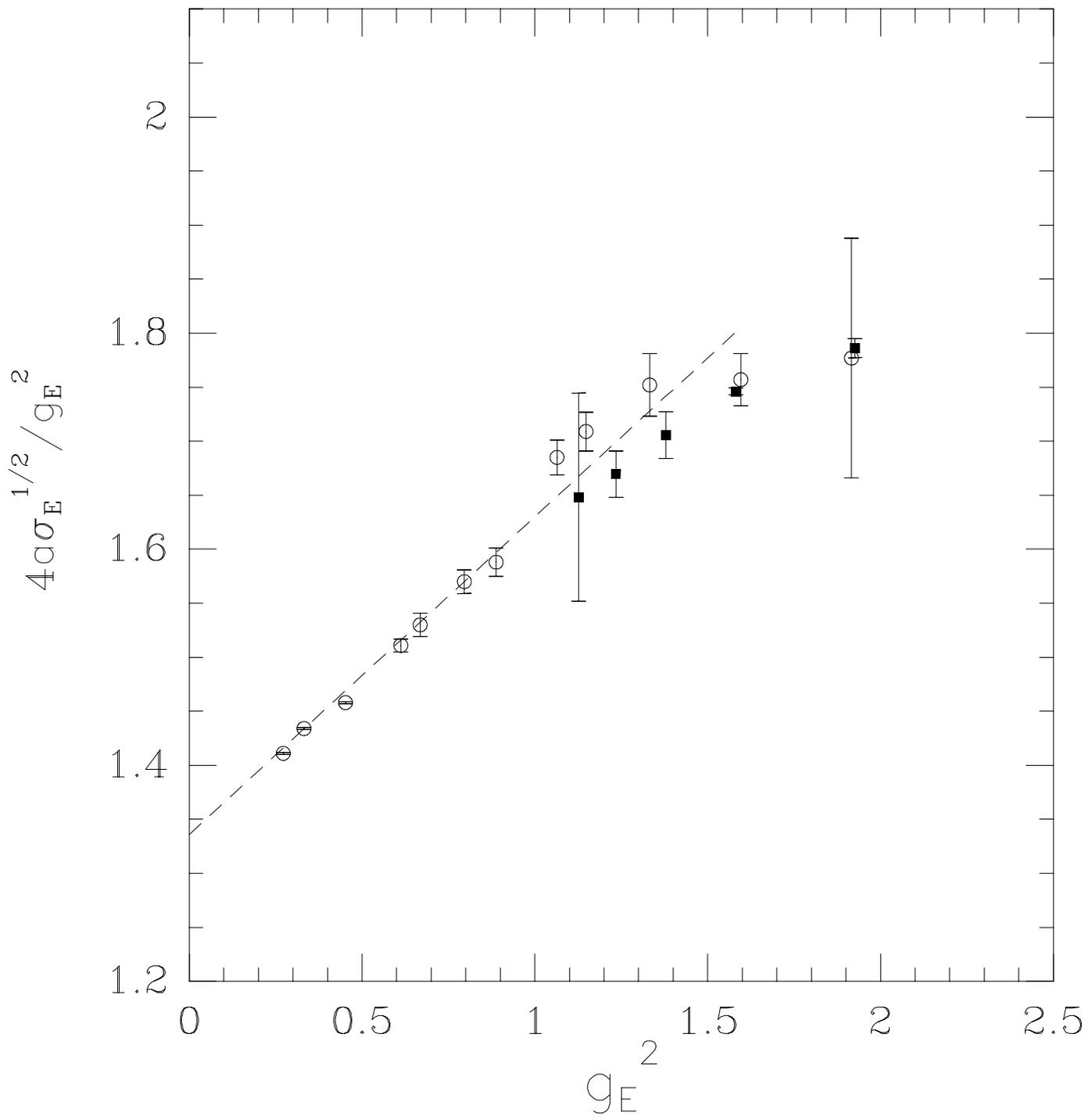
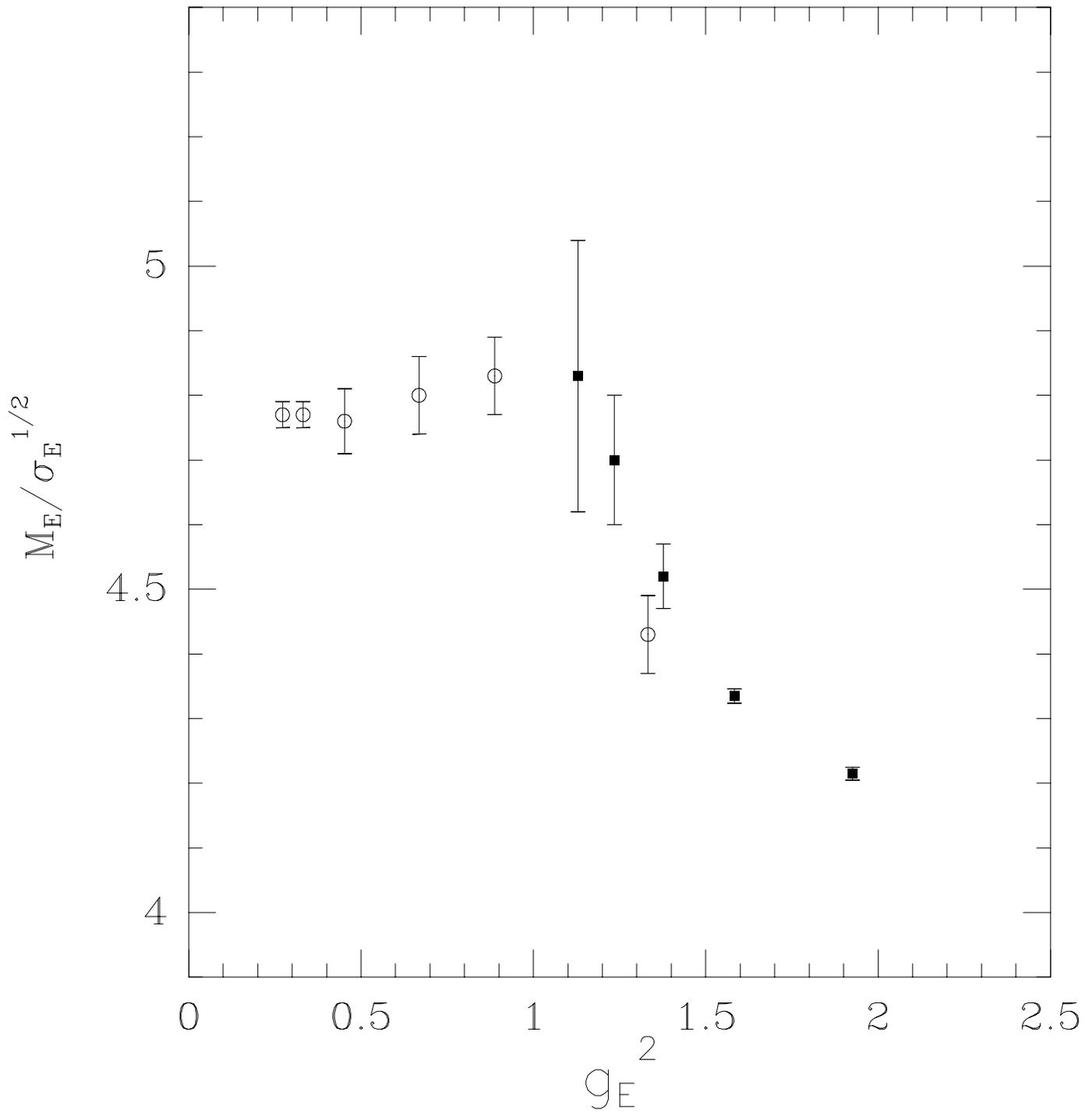
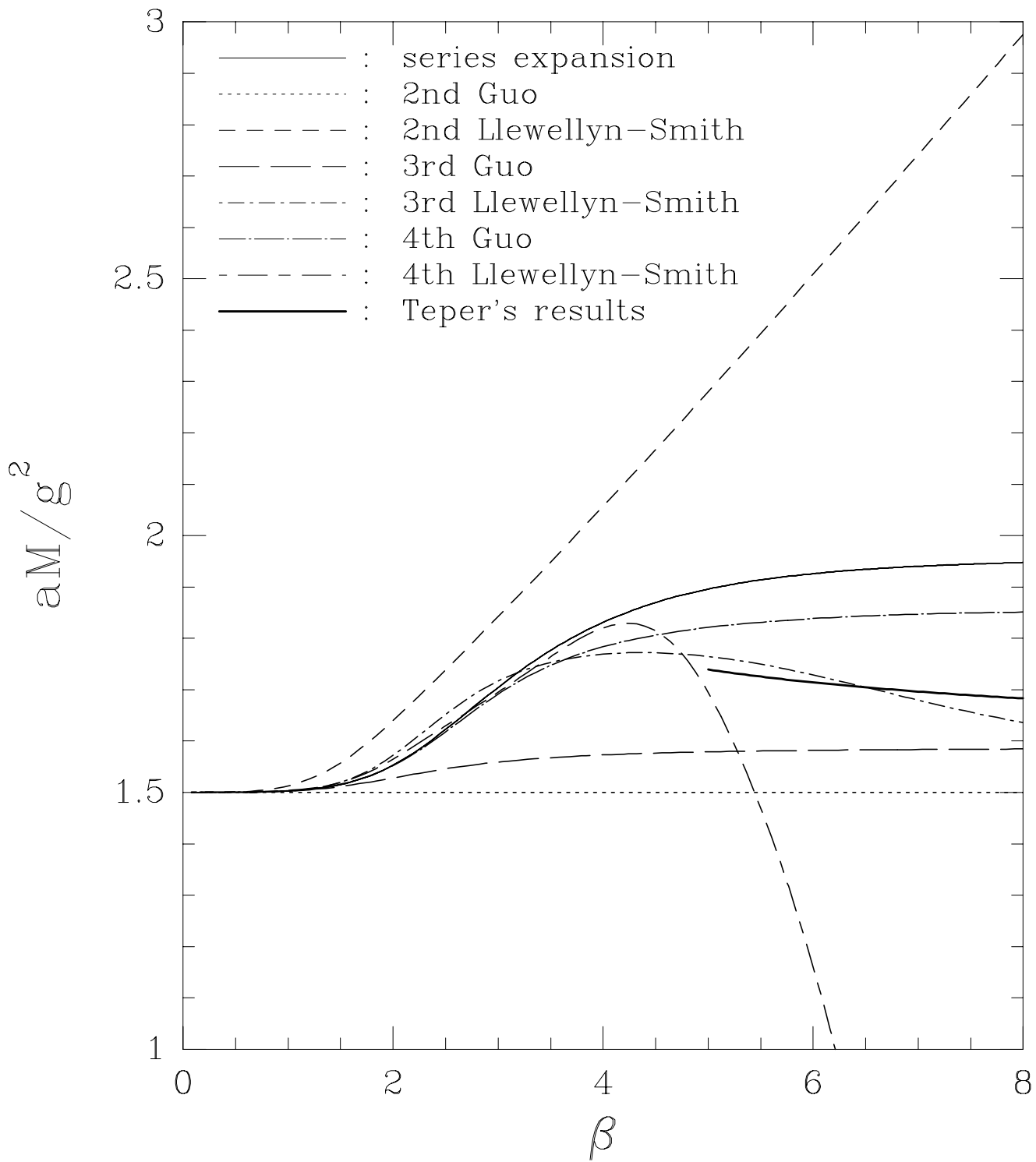
\end{document}